\documentclass[twocolumn,showpacs,preprintnumbers,amsmath,amssymb]{revtex4}
\usepackage{graphicx}
\usepackage{dcolumn}
\usepackage{bm}
\begin{document}
\preprint{APS/123-QED}
\title{EXPANSION OF A FORMALISM OF CLASSICAL MECHANICS
FOR NONEQUILIBRIUM SYSTEMS}
\author{V.M. Somsikov}
 \altaffiliation[] {}
 \email{nes@kaznet.kz}
\affiliation{%
Laboratory of Physics of the geoheliocosmic relation, Institute of
Ionosphere, Almaty, Kazakstan.
}%

\date{\today}
\begin{abstract}
The expansion of a classical Hamilton formalism consisting in
adaptation of it to describe the nonequilibrium systems is
offered. Expansion is obtained by construction of formalism on the
basis of the dynamics equation of the equilibrium subsystems by
which the nonequilibrium system is represented. It has allowed
removing restrictions on dynamics of the subsystems, which
dictated by the requirement of monogenic and potentiality of the
forces between subsystems. Modified Lagrange, Hamilton and
Liouville equations are obtained. Some features of dynamics of
nonequilibrium systems are considered. Connection between the
equation of interaction of subsystems and a thermodynamic
principle of energy is analyzed.
\end{abstract}

\maketitle

\section{\label{sec:level1}Introduction\protect}

The attempts of further development of methods of researches of
nonequilibrium systems collides with a problem of irreversibility
[1-4]. The analysis of approaches to solution of this problem leads
us to conclusion, that the one of the general reason of its
difficulties is a restriction on the frameworks of formalisms of the
classical mechanics. Really, creation of the organized structures is
caused by the dissipations. But dissipation is not present in
Hamilton systems because the obtaining of the Hamilton principle
based on condition that the forces of interaction of systems are
potentially and monogenic [5].

Necessity of expansion of formalisms of the classical mechanics have
arisen as a result of investigating of dynamics of hard disks [6,
7]. Forces of interaction of hard disks systems are non-potential.
Therefore it was necessary so to modify canonical Lagrange, Hamilton
and Liouville equations that they have been applicable for the
description of dynamics of non-potentially of interacting systems.
Such modification consisted in obtaining of these equations basing
on the D'Alambert principle under condition of non-potentiality of
the forces [5]. From modified Liouville equation followed that
non-potentiality of collective forces of interacting of systems is a
necessary condition of existence of irreversible dynamics [7].

In the nature all fundamental forces are potential. Therefore it was
necessary to generalize results of researches of hard disks on
systems of potentially interacting elements. For this purpose as a
method of the description of dynamics of nonequilibrium systems we
have taken advantage of an opportunity of representation of
nonequilibrium systems as a set of equilibrium subsystems (ESS).
This method of system's splitting on the ESS is not new. It was used
with success at construction of statistical physics of equilibrium
systems when interaction ESS can be neglected. But it follows from a
condition of maximum entropy that ESS for nonequilibrium system will
be in motion in relative to each other [8]. Therefore  to determine
their dynamics it is necessary to take into account the interaction
between them. The equation of interaction of systems (UVS) for
calculation of the ESS dynamics was obtained within the framework of
classical mechanics [9]. Using UVS and basing on D'Alambert
principle about equality to zero of a variation of work of effective
forces, the modified Lagrange, Hamilton and Liouville equations for
interaction ESS have obtained. These equations determine expansion
of a formalism of the classical mechanics, allowing study of the
open and nonequilibrium systems.

This paper is constructed as follows. The UVS and modified Lagrange,
Hamilton and Liouville equations are obtained. The analysis of the
important laws of dynamics of nonequilibrium systems and relation
between classical mechanics and thermodynamics are submitted.

\section{THE EQUATION OF ONE SYSTEM DYNAMICS}

Let us show, how obtaining of the equations of dynamics of
elements and their systems are possible. The equation of dynamics
of two interacting systems will be obtained in a similar way.

Let us consider system of potentially interacting elements. If the
system is conservative, we will have for energy $E$: $\dot{E}=0$.
The energy of the system depends on elements velocities, and their
coordinates. Therefore we have: $E=E(r,v)$, where $r,v$ are a set of
coordinates and velocities of the particles. In this case the
equality, $\dot{E}=0$, takes place only when energy depends on two
additive parts. One part should be a function of velocity, and the
second one - coordinates. Then energy of the system can be presented
as $E=\varphi{[T+U]}=const$, where $T=\sum\limits_{i=1}^{N}
T_i({v_i}^2)$, $v_i$ is a velocity of $i$- element, $T$ is the
kinetic energy of the system, and $U(r_i)$ is the potential one
[10]. The $\varphi$-function should be linear in order to be
constant when the coordinates and velocities are being changed. It
is always possible to represent such function as $E=T+U$ by means of
scale transformation and usage of the necessary system of
coordinates. So, the sum of kinetic and potential energies of the
system in a non-homogeneous space should be constant.

Let us take a moving elementary particle with mass $m$ and
velocity $v$. The kinetic energy corresponding to the particle
will be $T(v^2)=mv^2/2$, and the potential energy -$U(r)$, so
$E=m{v^2}/2+U(r)=const$. In this case from equality $\dot{E}=0$
follows that:
\begin{equation}
v(m\dot{v}+\partial{U}/\partial{r})=0\label{eqn1}
\end{equation}
The eq. (1) is a balance equation of the kinetic and potential
energies. As eq. (1) should be fulfilled in case of any direction of
the vector velocity, the following requirement takes place:
\begin{equation}
m\dot{v}=-\partial{U}/\partial{r}\label{eqn2}
\end{equation}
It is Newton equation (NE). This equation is determining the
connection of acceleration of the particle with the external force.
The right hand side of eq. (2) is the active force. The left hand
side - is the inertial force [5]. The particle moves along the
gradient of a potential function. During the motion of the particle
along the closed line in a potential field, the work of forces is
equal to zero. Therefore dynamics of the particle is reversible.

Let us take a system, which consists of $N$ potentially interacting
elements located in the non-homogeneous space; the mass of each
element is equal to 1. The force acting on each element is equal to
the sum of forces of all elements and the force caused by
non-homogeneous space. The force between two elements is central and
it depends on the distance between them.

Let us represent the energy of the system as a sum of kinetic
energies of elements - $T_N=\sum\limits_{i=1}^{N} m{v_i}^2/2$,
potential energies in the field of external forces - ${U_N}^{env}$
, and the potential energy of their interaction
${U_N}(r_{ij})={\sum\limits_{i=1}^{N-1}}{\sum\limits_{j=i+1}^{N}}U_{ij}(r_{ij})
$, where $v_i$ - is a velocity of $i$ -element; $r_{ij}=r_i-r_j$ -
is a distance between elements $i$ and $j$. So,
$E=E_N+U^{env}=T_N+U_N+U^{env}=const$. The time derivative of the
energy will be as follows: $\dot{E}=\dot{E}_N+\dot{U}^{env}$,
where
$\dot{E}_N={\sum\limits_{i=1}^{N}}v_i(m\dot{v}_i+\sum\limits_{j\neq{i}}^{N}F_{ij})$;
$\dot{U}_{env}=\sum\limits_{i=1}^{N}v_iF_i^{env}$;
$F_{ij}=\partial{U_N}/\partial{r_{ij}}$,
$F_{i}^{env}(r_i)=\partial{U^{env}}/\partial{r_{i}}$. So we have:
\begin{equation}
{\sum\limits_{i=1}^{N}}v_i\tilde{F}_{i}=0 \label{eqn3}
\end{equation}

Where $\tilde{F}_i=m\dot{v}_i+
\sum\limits_{j\neq{i}}^{N}F_{ij}+F_i^{env}$ is effective force for
$i$ particle. Then the eq. (3) can be rewritten as:
${\dot{E}=\sum\limits_{i=1}^{N}}v_i\tilde{F}_{i}=0$. This equality
can be treated as orthogonality of the vector of effective forces
with respect to the vector of velocities of elements of the
system. If there are no restrictions imposed on the $v_i$
directions, the requirement $\tilde{F}_i=0$ is satisfied [5]. Then
from eq. (3) we obtain:

\begin{equation}
{m\dot{v}_i=-\sum\limits_{i=1}^{N}}v_i\tilde{F}_{i}-F_i^{env}
\label{eqn4}
\end{equation}

It is NE for the system's elements in non-homogeneous space. The
eq. (4) has coincided with the corresponding equation in [11]
though it is obtained by other way.

Let us obtain an equation of motion of the system as a whole in an
external field. It is known [10], that a motion of the system in
the homogeneous space is characterized by two integrals: energy of
motion of the center of mass and the total energy of motion of the
particles inside of the system relative to the center of mass. We
will name second energy as the internal energy of the system.
Therefore it is reasonable to rewrite the eq. (3) having presented
systems energy as the total of these two types of energies. For
this purpose we can use the following equality:
$T_N=\sum\limits_{i=1}^{N}
m{v_i}^2/2=m/(2N)\{V_N^2+\sum\limits_{i=1}^{N-1}\sum\limits_{j=i+1}^{N}v_{ij}^2\}$
(a), where $V_N=\dot{R}_N=1/N\sum\limits_{i=1}^{N}\dot{r}_i$ -are
velocities of the center of mass; $R_N$ - are coordinates of the
center of mass; $v_{ij}=\dot{r}_{ij}$. We will write the energy of
the system in such a way: $E_N=T_N^{tr}+E_N^{ins}$,
$E_N^{ins}=T_N^{ins}+U_N$. Then the eq. (3) can be written as
follows:

\begin{equation}
\dot{T}_N^{tr}+ \dot{E}_N^{ins}
=-\sum\limits_{i=1}^{N}v_iF_i^{env}\label{eqn5}
\end{equation}
Where $\dot{T}_N^{tr}=M_NV_N\dot{V}_N; M_N=mN$;
$\dot{E}_N^{ins}=\dot{T}_N^{ins}+\dot{U}_N^{ins}$=
$\sum\limits_{i=1}^{N-1}\sum\limits_{j=i+1}^{N}v_{ij}(m\dot{v}_{ij}/N+F_{ij})$.

The term $\dot{T}_N^{tr}$ in eq. (5) determines the change of the
kinetic energy of motion of the system as a whole, and the term
$\dot{E}_N^{ins}$ represents change of the internal energy. The
right hand side determines the change of energy of the system as a
result of action of external forces.

Let us to transfer to the generalized variables in eq. (5). For
this purpose we will represent velocity of the motion of elements
of the system as the sum of velocities of their motion with
respect to the center of mass of the system - $\tilde{v}_i$ , and
velocity of the center of mass itself - $V_N$ , i.e.
$v_i=\tilde{v}_i+V_N$. Using these variables we will have the
following: $T_N=\sum\limits_{i=1}^{N}
m{v_i}^2/2=m/(2N)V_N^2+mV_N\sum\limits_{i=1}^{N}\tilde{v}_i
+\sum\limits_{i=1}^{N}m\tilde{v}_{i}^2/2$. As
$\sum\limits_{i=1}^{N}\tilde{v}_i=0$, then
$T_N=m/(2N)V_N^2+\sum\limits_{i=1}^{N}m\tilde{v}_{i}^2/2$.
Therefore $\sum\limits_{i=1}^{N}m\tilde{v}_{i}^2/2=
1/(2N)\sum\limits_{i=1}^{N-1}\sum\limits_{j=i+1}^{N}v_{ij}^2$.
Thus the kinetic energy of the relative motion of particles of the
system equals the sum of kinetic energies of the particles' motion
with respect to the center of mass.

Let us take into account that
$r_{ij}=\tilde{r}_{ij}=\tilde{r}_i-\tilde{r}_j$, where
$\tilde{r}_i, \tilde{r}_j$ - are coordinates of the elements with
respect to the center of mass of the system. So
$U_N(r_{ij})=U_N(\tilde{r}_{ij})=U_N(\tilde{r}_i)$. That is
$\sum\limits_{i=1}^{N-1}\sum\limits_{j=i+1}^{N}v_{ij}F_{ij}(r_{ij})
=\sum\limits_{i=1}^{N}\tilde{v}_iF_i(\tilde{r}_i)$, where
$F_i=\partial{U_N}/\partial{\tilde{r}_i}
=\sum\limits_{j\neq{i},j=1}^{N}\partial{U_N}/\partial{r_{ij}}$. By
means of generalization of corresponding equalities for the
kinetic and potential energies of the system, we will obtain from
eq. (5):
\begin{eqnarray}
V_NM_N\dot{V}_N+
\sum\limits_{i=1}^{N}m\tilde{v}_i(\dot{\tilde{v}}_i+F(\tilde{r})_i)=\nonumber\\=
-V_NF^{env}-\sum\limits_{i=1}^{N}\tilde{v}_iF_i^{env}(R,\tilde{r}_i)\label{eqn6}
\end{eqnarray}

Here $F^{env}=\sum\limits_{i=1}^{N}F_i^{env}(R,\tilde{r}_i)$,
$R$-is a coordinate of center of mass.

The eq. (6) determines the balance of energy of the system in
non-homogeneous space. The first term in the left hand side
corresponds to change of kinetic energy of motion of system as the
whole. The second term determines the change of internal energy.

Let us take into account that $F^{env}=F^{env}(R+\tilde{r}_i)$,
and suppose that $R\gg\tilde{r}_i$. Then it is possible to expand
the force $F^{env}$ in a series using a small parameter,
$\tilde{r}_i/R$. Keeping the terms up to first-order of
infinitesimal, we will have:
$F_i^{env}=F_i^{env}|_{R}+(\nabla{F_i^{env}})|_{R}\tilde{r}_i\equiv
F_{i0}^{env}+(\nabla{F_{i0}^{env}})\tilde{r}_i$. Taking into
account that $\sum\limits_{i=1}^{N}\tilde{v}_i
=\sum\limits_{i=1}^{N}\tilde{r}_i=0$ and
$\sum\limits_{i=1}^{N}F_{i0}^{env}=NF_{i0}^{env}=F_0^{env}$, we
can set from (6):
\begin{eqnarray}
V_N(M_N\dot{V}_N)+
\sum\limits_{i=1}^{N}m\tilde{v}_i(\dot{\tilde{v}}_i+F(\tilde{r})_i)\approx\nonumber\\\approx
-V_NF_0^{env}-({\nabla}F^{env}_{i0})\sum\limits_{i=1}^{N}\tilde{v}_i\tilde{r}_i\label{eqn7}
\end{eqnarray}

The force $F_0^{env}$ is potential and depends on $R$. The right
hand side of eq.(7) determines the work of external forces. The
first term is depending on the velocity of systems motion and
coordinate of its center of mass. It is determine the change of
kinetic energy of system as the whole.

The second term is depending on coordinates of particles and their
velocities in relative to the center of mass. It is determine the
change of internal energy of system. Thus, work of external forces
is splitting on two essentially different parts.

The right hand side is the first order of infinitesimal, because
despite $ R\gg\tilde{r}_i$, it value are not small.

If the external field is homogeneous then
$({\nabla}F^{env})_{i0}=0$. The variables in the eq. (7) is
separating, and we have:
\begin{equation}
M_N\dot{V}_N=F_0^{env}\label{eqn8}
\end{equation}
\begin{equation}
m\dot{\tilde{v}}_i=-F(\tilde{r_i})_i \label{eqn9}
\end{equation}

The eq. (8) is an equation of the motion of the system as a whole
in an external field, and the eq. (9) is an equation of the motion
of elements of the system. It follows from these equations, that
in a homogeneous external field the kinetic energy of the system
as a whole is variable only, and the internal energy is constant.
The change of this kinetic energy does not depend on the forces
between elements. The system is equivalent to the single body with
the mass which is equal to the sum of masses of the particles. The
motion of elements of the system does not depend on an external
field and is determined by forces of interaction only.

Thus, the system of potentially interacting elements in an external
field can be considered, us a structured particle. Generally the
work of external forces changes both its internal energy, and energy
of its movement. Therefore dynamics of such system (or the
structured particle) is determined by the equation (6). When the
heterogeneity of external forces can be neglected, dynamics of
system is described by the Newton equation.

\section{The equation of interaction of two subsystems}

Let us accept that the nonequilibrium system can be presented by
set of moving relative to each other of ESS. For ESS a role of
external forces carry out forces from other ESS. Obviously, it is
impossible to neglect by heterogeneity of these forces in
generally. Therefore dynamics ESS will be described by the
equation (6) in which the external forces is a forces from other
ESS. Hence, having replaced in (6) the right-hand side on these
forces, we will obtain the UVS which describe the dynamics of
nonequilibrium system.  Let us show, how it is possible to obtain
UVS.

Let us the system consists of two interacting equilibrium subsystems
- $L$ and $K$. Let us all elements to be identical and have the same
weight 1, and $L$ to be a number of elements in $L$ - subsystem, $K$
-is a number of elements in $K$ -subsystem, i.e. $L+K=N$,
$V_L=1/L\sum\limits_{i=1}^{L}v_i$ and
$V_K=1/K\sum\limits_{i=1}^{K}v_i$ - are subsystems' velocities with
respect to the center of mass. Let the velocity of the center of
mass be equal to zero, i.e. $LV_L+KV_K=0$.

We can represent the energy of the system as
$E_N=E_L+E_K+U^{int}=const$, where $E_L$ and $E_K$ are the
subsystems' energies, and $U^{int}$ - is the energy of their
interaction. According to the eq. (4), the energy of each
subsystem can be represented as $E_L=T_L^{tr}+E_L^{ins}$,
$E_K=T_K^{tr}+E_K^{ins}$, where $T_L^{tr}={M_L}V_L^2/2$,
$T_K^{tr}={M_K}V_K^2/2$, $M_L=mL, M_K=mK$. $E^{ins}$- is the
internal energy of a subsystem. The $E^{ins}$ consists of the
kinetic energy of motion of the elements with respect to the
center of mass - $T^{ins}$ and their potential energy - $U^{ins}$,
i.e. $E^{ins}=T^{ins}+U^{ins}$, where
$U_L^{ins}=\sum\limits_{i_L=1}^{L-1}\sum\limits_{j_L=i_L+1}^{L}U_{{i_L}{j_L}}(r_{i_Lj_L})$,
$U_K^{ins}=\sum\limits_{i_K=1}^{K-1}\sum\limits_{j_K=i_K+1}^{K}U_{{i_K}{j_K}}(r_{i_Kj_K})$.
The energy $U^{int}$ is determined as
$U^{int}=\sum\limits_{j_K=1}^{K}\sum\limits_{j_L=1}^{L}U_{j_Lj_K}(r_{j_Lj_K})$.
Indexes $j_k,j_L,i_K,i_L$ determine belonging of the elements to
corresponding subsystem. In equilibrium $T^{tr}=0$. Hence, if the
system aspirates to equilibrium, then $T^{tr}$ energy for each
subsystem will be transformed into the internal energy.

We have obtained the equations of dynamics of $L$ and $K$ ESS in
the following way. Let us differentiate energy of system on time.
In order to find the equation for $L$ - subsystem, at the left
hand side of obtained equality we have kept only those terms which
determine the change of the kinetic and potential energies of
interaction of elements of $L$ - subsystem. We replaced all other
terms in the right hand side and combined the groups of terms in
such a way when each group represents plurality of the terms
forming the NE for $K$-subsystems elements. Therefore these groups
are equal to zero. As a result the right hand side of the equation
will be contain only the terms which determine the change of the
potential energy of interaction of the elements $L$ - subsystem
with the elements $K$ - subsystem. The equation for $K$ -
subsystem we will obtain in the same way. So we will have:

${\sum\limits_{{i_L}=1}^L}mv_{i_L}{\dot{v}_{i_L}}+
{\sum\limits_{i_L=1}^{L-1}}{\sum\limits_{j_L=i_L+1}^{L}v_{{{i_L}{j_L}}}}{F_{{i_L}{j_L}}}
=-\sum\limits_{i_L=1}^{L}v_{{i_L}}F^K_{i_L}$ (b),

and for $K$-subsystem:

${\sum\limits_{{i_K}=1}^K}mv_{i_K}{\dot{v}_{i_K}}+
{\sum\limits_{i_K=1}^{K-1}}{\sum\limits_{j_K=i_K+1}^{K}v_{{{i_K}{j_K}}}}
F_{{i_K}{j_K}}$= ${\sum\limits_{j_K=1}^K}v_{{j_K}} F^L_{j_K}$ (c).

Here $F^K_{i_L}(R_K,r_{i_L})=\sum\limits_{{j_K}=1}^KF_{i_Lj_K}$,
$F^L_{j_K}(R_L,r_{i_K})=\sum\limits_{{i_L}=1}^LF_{i_Lj_K}$ - are
forces between the corresponding particle of one ESS and all
particles of the other ESS. The work of these forces determines the
change of energy of ESS. $R_K=(1/K)\sum\limits_{{i_K}=1}^Kr_{i_K}$;
$R_L=(1/L)\sum\limits_{{i_L}=1}^Lr_{i_L}$.

For obtaining from the equations (b, c) the equations of subsystems
interaction, it is necessary to present their left-hand side in the
form of two terms. One term should characterize the change of energy
of a motion of a ESS as the whole, and second is a change of its
internal energy. With this purpose let us take into account that
velocities of the elements can be represented as the sum of their
velocities with respect to the center of mass of the ESS and the
velocity of the ESS with respect to the center of mass of the
system, $ v_i=\tilde{v}_i+V$. Then with a help of (a), we will
obtain from (b, c):
\begin{eqnarray}
V_LM_L\dot{V}_L+{\sum\limits_{i_L=1}^{L-1}}\sum\limits_{j_L=i_L+1}^{L}\{v_{i_Lj_L}
[\frac{{m\dot{v}}_{i_Lj_L}}{L}+\nonumber\\+F_{i_Lj_L}]\}=-{\Phi}_L-V_L{\Psi}
\end{eqnarray}
\begin{eqnarray}
V_KM_K\dot{V}_K+{\sum\limits_{i_K=1}^{K-1}}\sum\limits_{j_K=i_K+1}^{K}\{v_{i_Kj_K}
[\frac{{m\dot{v}}_{i_Kj_K}}{K}+\nonumber\\+F_{i_Kj_K}]\}={\Phi}_K+V_K{\Psi}
\end{eqnarray}
Here $\Psi=\sum\limits_{{i_L}=1}^LF^K_{i_L}$,
 ${\Phi}_L=\sum\limits_{{i_L}=1}^L\tilde{v}_{i_L}F^K_{i_L}$,
 ${\Phi}_K=\sum\limits_{{i_K}=1}^K\tilde{v}_{i_K}F^L_{i_K}$.

The eqs. (10, 11) are equations of systems interaction (UVS). The
first terms in the left hand side of UVS is a change of motion
energy of ESS - "${\dot{T}}^{tr}$", and the second terms is a
change of their internal energy - "${\dot{E}}^{ins}$".

The right hand side of UVS are consisting of two terms determining
the work of collective forces of interaction ESS. The first terms,
"${\Phi}_L$ " and "${\Phi}_K$", determine the work of the
interaction force on transformation of motion energy of ESS into
their internal energy. The second terms include potential force,
"$\Psi$". It is a central force which depends on distances between
the centers of mass of ESS: $\Psi={\Psi}(R_{LK})$, where
$R_{LK}=R_L-R_K$. The work of "$\Psi$" determines the change of
velocities of ESS. As well as in the case of the motion of the
system in a non-homogeneous external field (see the eq. (7)), when
$R_L\gg\tilde{r}_{i_L}$, $R_K\gg\tilde{r}_{i_K}$ this work can be
estimated by expanding of force in a Taylor series.

It is possible to transform the equations (10, 11) to the form of
the equations containing forces of friction. Let us take into
account that
$\sum\limits_{{i_L}=1}^Lm\tilde{v}_{i_l}\dot{\tilde{v}}_{i_L}=
(1/L){\sum\limits_{i_L=1}^{L-1}}\sum\limits_{j_L=i_L+1}^{L}
mv_{{i_L}{j_L}}{\dot{v}}_{{i_L}{j_L}}$,
${\dot{U}}_L=\sum\limits_{{i_L}=1}^LF_{i_L}\tilde{v}_{i_L}=
{\sum\limits_{i_L=1}^{L-1}}\sum\limits_{j_L=i_L+1}^{L}F_{i_Lj_L}{\tilde{v}}_{i_Lj_L}$,
$F_{i_L}=\sum\limits_{{j_L}{{\neq}i_L}}^{L}{\partial}U_L/{\partial}
\tilde{r}_{i_L}$ . And for $K$ -subsystem is similar. If also
introduce the following designation:
${\alpha}_{L}=-(\dot{E}^{ins}_{L}+{\Phi}_{L})/V^2_{L}$,
${\alpha}_{K}=-(\dot{E}^{ins}_{K}-{\Phi}_{K})/V^2_{K}$,
${\dot{E}}_L^{ins}=\sum\limits_{{i_L}=1}^L{\tilde{v}}_{i_L}
(m{\dot{\tilde{v}}}_{i_L}+F_{i_L})$,
${\dot{E}}_K^{ins}=\sum\limits_{{i_K}=1}^K{\tilde{v}}_{i_K}
(m{\dot{\tilde{v}}}_{i_K}+F_{i_K})$ then the eqs. (10,11) can be
rewritten as:
\begin{equation}
M_L\dot{V}_L=-\Psi-{\alpha}_LV_L \label{eqn12}
\end{equation}
\begin{equation}
M_K\dot{V}_K=-\Psi- {\alpha}_KV_K\label{eqn13}
\end{equation}

Here coefficients "$\alpha_L$", "$\alpha_K$" are equal to the
ratio between the energy of motion and the full kinetic energy of
correspondent ESS. Thus, the forces of interaction of the ESS are
divided into the potential and non-potential parts, and $\alpha$
is a similar to the friction coefficient. The non-potential part
of the force determined by coefficient corresponds to dissipation.
It is important to note that here energy is not disappeared.
Actually the value of energy is constant, but energy is
redistributed between the elements of the subsystem. The eqs. (12,
13) are equivalent to the empirical motion equations for the
structured particles.

Let us perform the following replacement: $V_L=KV_{LK}/N$,
$V_K=-LV_{LK}/N$, where $V_{LK}=\dot{R}_{LK}{\equiv}V_L-V_K$. As the
center of mass of the system is immovable, we will have from
eqs.(10, 11):
\begin{equation}
{\dot{E}}^{tr}_{LK}+{\dot{E}}^{ins}_{L}+{\dot{E}}^{ins}_{K}=\Phi
\label{eqn14}
\end{equation}
Here $\Phi=-\Phi_L+\Phi_K$,
${\dot{E}}^{tr}_{LK}=V_{LK}(M_R\dot{V}_{LK}+\Psi)$ - is the change
of energy of the relative motion of the ESS, $M_R=mKL/N$.

The second and third terms in the eq. (14) determine the change of
their internal energies. The right hand side determines the change
of interaction energy. It is obvious that for an equilibrium system
${\dot{E}}^{tr}_{LK}=0$. Then the eq. (14) is splitting on two
independent equations for ESS.

We see that the work of the forces between ESS goes not only on
change of their kinetic and potential energy, as in the case of
elementary particles. It also goes on change of their internal
energy. Therefore the force of interaction ESS represents the sum
of two various forces. The first force determines motion of ESS as
the whole as a result of mutual transformation their kinetic and
potential energies. It is potential force. The second force
determines the work on transformation of energy of interaction to
the internal energy of ESS. It is non-potential force. This force
does not change the momentums of ESS.

Thus, it is possible to consider equilibrium subsystems $L$ and
$K$ as the structured particles. The motion of ESS are determined
not only by transformations their kinetic and potential energy.
But it is determined by the changes of their internal energy olso.
Therefore the force of interaction ESS represents the sum of two
various forces. The first force determines motion of ESS as the
whole as a result of mutual transformation their kinetic and
potential energies. It is potential force. The second force
determines the work on transformation of motion energy into the
internal energy of ESS. It is non-potential force. This force does
not change the momentums of ESS.

If $L=1$, $K=N-1$ the equation (10) for $L$ - subsystem will
coincide with NE. The internal energy will be absent, and force
acting on it will be potential. Therefore its dynamics is
reversible. Therefore reversibility of dynamics of separate elements
 does not mean yet reversibility of dynamics of their systems. It is
essential difference UVS from NE for elementary particles because
in according with NE, the elementary particles return into initial
points if the direction of velocity of their centers of mass will
change on opposite, but ESS does not come back into initial points
of their phase space when reversing them velocities.

Generally the potentials of interaction between particles inside ESS
can be distinct from potentials of interaction of particles for
different ESS.

\section{The Lagrange, Hamilton and Liouville modified equations}

There is a question as in connection with the UVS the canonical
Lagrange, Hamilton and Liouville are modified. The canonical
equations follow from the integral principle of Hamilton [5, 14]. In
turn the integral principle of Hamilton follows from differential
principle of D'Alambert in case of potentiality of active forces
between ESS. D'Alambert equation is formulated on the basis of NE.

In accordance with D'Alambert principle: "the work of the effective
forces which includes the inertial and active forces is equal to
zero for all reversible virtual displacements compatible with the
restrictions given" [5]. For the closed non-equilibrium system that
represents themselves by a set of the ESS, this principle will be as
follows: the work of all forces of interaction of the ESS is equal
to zero for all reversible virtual displacements compatible with the
restrictions given.

If the change of the internal energy of ESS can be neglected, the
work on their motion will be determined only by the potential part
of interaction ESS forces. In this case instead UVS we can use NE
and therefore come to the known canonical equations of a classical
mechanics [5]. If the neglecting by the change of internal energy is
impossible, D'Alambert equation should be written down on the basis
UVS which take into account the distribution of the subsystems'
motion energy as a whole between its internal energies in the result
of the work of the non-potential part of forces.

Let us take a closed non-equilibrium system consisting of $N$
elements which can be represented by a set of $R=2$ an ESS. The
modification of formalism is carried out by standard way [5].
Firstly, on the basis of UVS the D'Alambert equation is obtained.
Based on it a principle of Hamilton, and then Lagrange equation are
obtained. After that the Hamilton and Liouville equations for ESS
are deduced.

If there are no external restrictions on system then the virtual
displacements of the elements and ESS can be combined with the real
ones [5]. As virtual work of all forces is determined by the eq.
(14) we will have:
\begin{equation}
\delta\bar{W}_{LK}={\delta}E^{tr}_{LK}+ {\delta}E^{ins}_{L}
+{\delta}E^{ins}_K +{\delta}\Phi=0 \label{eqn15}
\end{equation}
The eq. (15) is D'Alambert equation for a non-equilibrium system.
The first three terms in (15) determine the virtual change of
energies of the ESS. The two last terms determine the virtual work
for displacement of the elements of one ESS in the field of forces
of the elements of the other ESS. I.e. they determine the change
of internal energies of the ESS. The line above
$\delta\bar{W}_{LK}$ means that it is just a differential form
which does not come to a scalar function variation.

Let us transform the eq. (15) by means of multiplying by $dt$, and
integrating from $t=t_1$ to $t=t_2$. As a result we will have:
\begin{eqnarray}
\int_1^2{\delta\bar{W}_{LK}}dt=\int_1^2[{\delta}E^{tr}_{LK}+
{\delta}E^{ins}_{L} +{\delta}E^{ins}_K+\nonumber\\
+\sum\limits_{i_L=1}^{L}{\delta}{\tilde{r}}_{i_L}F_{i_L}-
\sum\limits_{i_K=1}^{K}{\delta}{\tilde{r}}_{i_K}F_{i_K}]dt=0
\label{eqn16}
\end{eqnarray}
where ${\delta}{\tilde{r}}_{i_L}$ -is a virtual displacements of
$i_L$ element of $L$ -subsystem; ${\delta}{\tilde{r}}_{i_K}$ - is
a virtual displacements of $i_K$  element of $K$- subsystem.

Let us rewrite the first three terms as follows:
${\delta}{\int}_1^2[\sum\limits_{i_L=1}^{L}M_RV^2_{LK}/2+U_{LK}]dt-
[\sum\limits_{i_L=1}^{L}M_RV_{LK}{\delta}R_{LK}]_{t_1}^{t_2}$;
${\delta}{\int}_1^2[\sum\limits_{i_L=1}^{L}m{\tilde{v}}_{i_L}^2/2+U_L]dt-
[\sum\limits_{i_L=1}^{L}m{\tilde{v}}_{i_L}{\delta}r_{i_L}]_{t_1}^{t_2}$;
${\delta}{\int}_1^2[\sum\limits_{i_K=1}^{K}m{\tilde{v}}_{i_K}^2/2+U_K]dt-
[\sum\limits_{i_K=1}^{K}m{\tilde{v}}_{i_K}{\delta}r_{i_K}]_{t_1}^{t_2}$;

The last two items in (16) are not the full differentials from any
scalar function. I.e. in general it is impossible to represent the
active forces of ESS interaction in the form of a gradient from a
forcing function. Such forces are named as polygenic ones.

Let us the virtual displacements be equal to zero at the ends of
the interval $[t_1, t_2]$. Then the eq. (16) can be rewritten as
follows:
\begin{eqnarray}
{\delta}S=
{\int}_1^2{\delta}{\bar{W}}_{LK}dt={{\int}_1^2}{\{}\sum\limits_{n=1}^{N}
[d/dt({\partial}{\Im}/{\partial}v_n)-\nonumber\\-{\partial}{\Im}/{\partial}r_n+F_n]{\}}{\delta}r_ndt
\label{eqn17}
\end{eqnarray}

Here $S$- is action, $\Im=\sum\limits_{n=1}^{N}[mv_n^2+U_n]$ - is
Lagrangian, $F_n=F_n^L-F_n^K$.

All variables are independent, and that is why in order to achieve
simplification we have used through indexing for all elements of
the system instead of indexing in accordance with the ESS, and at
the same time new indexing is in accordance with the one accepted
in the eq. (16). Thus, the action for a non-equilibrium system
consists of the terms describing dynamics of the ESS as a whole,
interior dynamics of the elements of the ESS and the term
describing interaction of the ESS.

It follows from the requirement of independence of the generalized
variables that
\begin{equation}
d/dt({\partial}{\Im}/{\partial}v_n)-{\partial}{\Im}/{\partial}r_n=-F_n
\label{eqn18}
\end{equation}
This is the very modified Lagrange equation for a non-equilibrium
system. The right hand side of the eq. (18) is equal to zero only
when the relative motion of the ESS are absent. It should take
place when the system is in equilibrium.

The eqs. (10, 11) were obtained by means of independent methods
from the requirement of the system energy to be constant. It
allows us to write the modified Lagrange equation for each of the
ESS at once.

\begin{equation}
d/dt({\partial}{\Im}_L/{\partial}v_{i_L})-{\partial}{\Im}_L/{\partial}r_{i_L}=-F_{i_L}^K
\label{eqn19}
\end{equation}

\begin{equation}
d/dt({\partial}{\Im}_K/{\partial}v_{i_K})-{\partial}{\Im}_K/{\partial}r_{i_K}=-F_{i_K}^L
\label{eqn20}
\end{equation}

Here ${\Im}_L$ and ${\Im}_K$ - are Lagrange functions for the ESS.

Now let's obtain modified Hamilton equation. For $\Im$ the
differential can be written as $d\Im$=
$\sum\limits_{n=1}^{N}[({\partial}{\Im}/{\partial}v_{n})dv_n$+
$({\partial}{\Im}/{\partial}r_ndr_n)]+{\partial}{\Im}/{\partial}tdt$.
With the help of Legendry transformation, we will obtain:
$d[\sum\limits_{n=1}^{N}p_nv_n-\Im]=\sum\limits_{n=1}^{N}[v_ndp_n-
({\partial}{\Im}/{\partial}r_n)dr_n]-({\partial}{\Im}/{\partial}t)dt$.
Since ${\partial}{\Im}/{\partial}t$=$-{\partial}{H}/{\partial}t$ ,
where $H=d[\sum\limits_{n=1}^{N}p_nv_n-\Im]$, we will have:

\begin{equation}
{\partial}{H}/{\partial}r_{n}=-\dot{p}_n-F_n
 \label{eqn21}
\end{equation}

\begin{equation}
{\partial}{H}/{\partial}p_{n}=v_n \label{eqn22}
\end{equation}

The above are modified Hamilton equations for a non-equilibrium
system. The right hand side of the eq. (21) denotes non-potential
forces. Modified Hamilton's equations for the ESS can be obtained
in the same way.

Now let us obtain Liouville equations for the ESS and full system.
For this porpoise let us take a generalized current vector
$J_L=J_L(v_n,{\dot{p}}_n)$  of the $L$-subsystem (here "$n$" index
corresponds only to the particles of the $L$- subsystem). From the
eqs. (21, 22) we find:
$divJ_L=\sum\limits_{n=1}^{L}({\partial}v_n/{\partial}r_n+
{\partial}{\dot{p}}_n/{\partial}p_n)=-
\sum\limits_{n=1}^{L}{\partial}F_n^K/{\partial}p_n$. The
continuity equation is a differential form of the particle number
conservation law: ${\partial}f_L/{\partial}t+div(f_LJ_L)=0$, where
$f_L=f_L(r,p,t)$ is a normalized distribution function for the
elements of the $L$- subsystem. With the help of the continuity
equation we will have:
$df_L/dt={\partial}f_L/{\partial}t+\sum\limits_{n=1}^{L}(v_n{\partial}f_L/{\partial}r_n+
\dot{p}_n{\partial}{f}_L/{\partial}p_n)$=${\partial}f_L/{\partial}t+div(f_LJ_L)-f_LdivJ_L=
-f_LdivJ_L$=$f_L\sum\limits_{n=1}^{L}{\partial}F_n^K/{\partial}p_n$.
Thus we have:
\begin{equation}
df_L/dt=f_L\sum\limits_{n=1}^{L}{\partial}F_n^L/{\partial}p_n
\label{eqn23}
\end{equation}
The eq. (23) is modified Liouville equation for the $L$
-subsystem.

There is a formal solution of the eq. (23):
$f_L={const}{\times}{\exp}{\int}{\{}\sum\limits_{n=1}^{L}{\partial}F_n^K/{\partial}p_n{\}}dt$.
From this solution follows that the distribution function of a
subsystem depends on time if force acting on the subsystem depends
on velocities of the elements of the subsystem. It is similar result
as for a hard disks systems [7].

The modify Liouville equation for system can be obtain from eqs.
(21, 22) if $n =1,2...N$. We will have:
\begin{equation}
df_N/dt=f_N{\partial}F_n/{\partial}p_n \label{eqn24}
\end{equation}
Despite absence of external forces for a system, the right hand side
of the eq. (24) is not equal to zero. It is caused by the fact that
the energy of subsystems' interaction does not include into ESS
energy. Therefore the right hand side of the eq. (24) will not be
equal to zero until the energy of interaction of the ESS is not
transformed into their internal energy.

The reason of distinction between two descriptions of system (in a
framework of canonical and modified equation) is the total work of
potential forces between particles in the closed system is equal
to zero. But the work on moving ESS is not equal to zero. The
Newton equation for select particles can not determine this work.
Therefore modified Lagrange, Hamilton and Liouville equations as
against their canonical prototypes, are applicable for the
description of dynamics of nonequilibrium systems.

\section{Some properties of dynamics of interacting subsystems}

In accordance with UVS, in order to describe a non-equilibrium
system, the energy should be divided into $T^{tr}$ and $E^{ins}$.
As UVS includes the parameter describing the degree of
non-equilibrium of a system, it allows studying evolution of the
system. Let us to analyze some important cases of the system's
dynamics.

In general case the non-potential forces can be rewritten as:
$\Phi_L=\sum\limits_{{i_L}=1}^L{\Phi}_{i_L}(R_K,\tilde{r}_{i_L},
\tilde{v}_{i_L})$,
$\Phi_K=\sum\limits_{{i_K}=1}^K{\Phi}_{i_K}(R_L,\tilde{r}_{i_K},
\tilde{v}_{i_K})$. If the sizes of ESS are less than the distances
between them, we will have
$\Phi_L\approx\sum\limits_{{i_L}=1}^LF^K_{i_L}(R_K)\tilde{v}_{i_L}$=
$F^K_{i_L}\sum\limits_{{i_L}=1}^L\tilde{v}_{i_L}=0$, and
$\Phi_K\approx\sum\limits_{{i_K}=1}^KF^L_{i_K}(R_L)\tilde{v}_{i_K}$=
$F^L_{i_K}\sum\limits_{{i_K}=1}^K\tilde{v}_{i_K}=0$. In this case
the eqs. (10, 11) are the Newton equations for two hard bodies. This
case is similar to the case of homogeneity of an external field of
forces.

In the case when $\Psi=0$, $V_L=V_K=0$, the first terms of the left
side of UVS are equal to zero and the system is in equilibrium. The
full energy of the system is equal to the sum of $E^{ins}$ for each
subsystem. Such system can be studied with the help of canonical
Hamilton equations. ESS. Gibbs used this approach for creation of
statistical mechanics of equilibrium systems [3, 11].

Now let us velocities of the elements with respect to the centers
of mass of corresponding ESS be equal to zero during interaction
of the ESS (this is a hard body approach). Then the right hand
side, and also the second terms of left hand side of UVS are equal
to zero, and UVS is transformed into NE for two hard bodies.

It is possible to come from classical mechanics to thermodynamics
with the help of the UVS. Really, the right hand side of eqs. (10,
11) determines an change of ESS energy. The first term of the left
hand side of UVS determines the change of the motion energy of ESS
as the whole. In thermodynamics it corresponds to work, which is
carried out by external forces acting on ESS on the part of an
environment. The second term of the left hand side corresponds to
increase in the entrance energy of a ESS. In thermodynamics this
term corresponds to the change of thermal energy of ESS.

Let us consider the relation between the UVS and the basic equation
of thermodynamics in details. We can write the basic equation of
thermodynamics [8,13]: ${dE=dQ-{PdY}}$. Here, according to common
terminology, $E$ is energy of a ESS; $Q$ is thermal energy; $P$ is
pressure; $Y$ is volume. We take into account that $N>>1, M_N=1$.
The energy change of the selected ESS is due to the work made by
external forces. Therefore, the change in full energy of a subsystem
corresponds to $dE$.

The change of kinetic energy of motion of a ESS as the whole,
$dT^{tr}$, corresponds to the term ${PdY}$. Really,
${dT^{tr}=VdV=V\dot{V}dt=\dot{V}dr=PdY}$

Let us determine, what term in UVS corresponds to the change of
the internal energy of ESS. As follows from virial theorem [10],
if the potential energy is a homogeneous function of second order
of the radiuses-vectors, then
${\bar{E}^{ins}=2\bar{\tilde{T}}^{ins}=2\bar{\tilde{U}}^{ins}}$.
The line denotes the time average. Earlier we obtained that the
energy, ${E^{ins}}$, increases due to ${T^{tr}}$. But the opposite
process is impossible. Therefore the change of the term $Q$ in the
UVS corresponds to the change of the energy ${E^{ins}}$.

Let us consider the ESS near to equilibrium. If the ESS consist of
${N}$ elements, the average energy of each element becomes,
${\bar{E}^{ins}={E}^{ins}/N=\kappa{T}_0^{ins}}$. Now let the
internal energy increases with ${dQ}$. According to the virial
theorem, keeping the terms of the first order, we have:
${dQ\approx{T}_0^{ins}[d{E}^{ins}/{T}_0^{ins}]
={T}_0^{ins}[{dv}/{v_0}]}$, where ${v_0}$ is the average velocity
of an element, and ${dv}$ is its change. For ESS in equilibrium,
we have ${dv/v_0\sim{{d\Gamma}/{\Gamma}}}$, where ${\Gamma}$ is
the phase volume of a ESS, ${d\Gamma}$ will increase due to
increasing of the ESS energy on the value, ${dQ}$. By keeping the
terms of the first order we get:
${dQ\approx{T}_0^{ins}d\Gamma/\Gamma={{T}_0^{ins}}d\ln{\Gamma}}$.
By definition ${d\ln{\Gamma}=dS^{ins}}$, where ${S^{ins}}$ is a
subsystem entropy [9]. So, near equilibrium we have
${dQ\approx{T}_0^{ins}dS^{ins}}$. So formally according with UVS
the entropy production in the non-equilibrium system is determined
by transformation of the motion energy of ESS into the internal
energy. If it so eventually relative velocities of ESS go to zero.
In result the energy of relative motion of ESS will be completely
transforms into the internal energy and the systems equilibrates.
It means that energy of motion of a ESS goes on increase of
entropy. Therefore the deviation of entropy from equilibrium can
be determined by the next formula [7]:
\begin{equation}
{{\Delta{S}}={\sum\limits_{l=1}^R{\{{m_l}
\sum\limits_{k=1}^{m_l}\int{\sum\limits_s{{\frac{{F_{ks}}^{m_l}v_k}{E^{m_l}}}}{dt}}\}}}}\label{eqn25}
\end{equation}

Here ${E^{m_l}}$ is the kinetic energy of ESS; ${m_l}$ is the number
elements in ESS ${"l"}$; ${R}$ is the number of ESS; ${s}$ is number
of the external element which collided with internal element ${k}$;
${F_{ks}^{m_l}}$ is a force, acted on element $k$-element; $v_k$ -is
a velocity of the $k$- element.

The formula (25) means, that a entropy production is provided by
the energy of relative motion of ESS. It corresponds to entropy
definition for the non-equilibrium systems, which was offered in
[8].

But for informal connection of the equations of interaction of
systems with thermodynamics, the increase of internal energy due to
the energy of relative motion should be irreversible. Below some
arguments for the benefit of such increase will be offered.

In connection with the law of momentum preservation of ESS the
internal energy of the ESS is gradually increased due to energy
${\dot{E}}^{tr}_{LK}$. I.e. there is a gradual decrease of
$V_{LK}$. To explain this fact let us take an equilibrium system
and divide it into ESS. It follows from the requirement of
equilibrium that the forces between the ESS are equal to zero. But
if there are no external forces with respect to a subsystem, the
subsystem cannot start moving. Indeed in this case the subsystem
can start moving only due to its internal energy. But it is
prohibited by the law of conservation of momentum of the
subsystem. I.e. if the system is in equilibrium then it will be
always remain in equilibrium state.

Now let us consider two interacting ESS which are in relative
motion. The relative velocity of ESS can not be increased due to
the energy of interaction of the ESS as this energy is itself
determined by the relative velocities of the ESS. As well as in
the previous case, this velocity can not be increased due to the
internal energy of the ESS. So relative velocity can be decreased
only.

It is follows from these arguments that UVS characterizes motion of
system to an equilibrium state as a result of transition of energy
of relative motion of ESS into internal energy. When system is in
equilibrium, or when it is possible to freeze internal degrees of
freedom of ESS, i.e. to neglect the change of their internal energy,
then UVS is reduced to NE. Hence, it is possible to assert, that the
broken of the time-symmetry of the systems dynamics is caused by the
increasing of the internal energy. The decreasing of internal energy
is prohibited by the law of conservation of momentum of the
subsystem.

\section{Conclusion}
Expansion of a formalism of classical mechanics and approach to the
analysis of nonequilibrium systems are based on the key statement
that closed non-equilibrium systems can be represented as a set of
moving ESS [8]. This statement allowed transforming the tasks of
dynamics of a non-equilibrium system to the tasks of dynamics of
ESS. The offered approach has allowed removed the restrictions of
applicability formalism of the classical mechanics, bound with the
requirement of conservatism of systems. It has appeared enough for
elimination of contradictions between a classical mechanics and
thermodynamics.

Having expressed energy of system in the form of energy of a
motion of ESS as the whole, their internal energies and energy of
their interaction and differentiated it on the time, the UVS has
been obtained. This equation is equivalent to the equation of
structured particles.

In agreement with UVS the dynamics of ESS is determined by the
collective force of their interaction. This force consists of two
parts. The potential part is determining the transformation of the
interaction energy into the motion energy of ESS. Non-potential
force is determining the transformation of the ESS motion energy
into internal energy. Thus, unlike dynamics of an elementary
particle, the dynamics of ESS is determined by the work of
interaction forces which will transform energy of interaction not
only into potential and kinetic energy of ESS motion as the whole,
how we have in the case of elementary particle, but also into the
internal energy. Therefore based on NE which is valid only for
elementary particles, it is impossible to describe the dynamics of
the nonequilibrium system.

Based on UVS the modified Lagrange, Hamilton and Liouville
equations was obtained. These equations are applicable to the
study of dynamics of the open non-equilibrium systems because its
have allowed taking into account the work of non-potential part of
force on transformation of energy of ESS motion into their
internal energy.

The following mechanism of irreversibility can be proposed: energy
of ESS relative motion is transformed into their internal energy as
a result of the work of the non-potential part of force of
interaction of ESS. The decreasing of internal energy is prohibited
by the law of conservation of momentum of the ESS. Therefore the
system equilibrates, when relative motion of ESS is absent.

The explanation of the First law of thermodynamics is based on the
fact that the work of subsystems' interaction forces changes both
the energy of their motion and their internal energy. The
explanation of the Second law of thermodynamics is based on the
condition of irreversible transformation of the subsystems' relative
motion energy into their internal energy.

The solution many problems are necessary for the further development
of the offered approach to studying nonequilibrium systems and the
strict proof of the conclusions offered here. Questions about the
ways of splitting of nonequilibrium system on ESS, about character
of transformation of energy of relative motion of ESS in their
internal energy, etc., are the most important between them.

\smallskip

\end{document}